%% file: main.tex
\documentclass[sigconf, authorversion]{acmart}

\AtBeginDocument{%
  \providecommand\BibTeX{{%
    \normalfont B\kern-0.5em{\scshape i\kern-0.25em b}\kern-0.8em\TeX}}}

\setcopyright{none}
\renewcommand\footnotetextcopyrightpermission[1]{} 
\usepackage{comment}
\usepackage{amsmath}
\usepackage{graphics}
\usepackage{color}
\usepackage{fontawesome}
\usepackage{todonotes}
\usepackage{xspace}
\usepackage{soul}
\usepackage{siunitx}
\usepackage[flushleft]{threeparttable}
\usepackage[title]{appendix}
\usepackage{enumitem}
\usepackage{balance}
\usepackage{caption}
\usepackage{graphics}
\usepackage{xspace}
\usepackage{color, colortbl, soul, xcolor}
\usepackage{tikz}
\usepackage{caption,subcaption}
\usepackage{fixltx2e}
\usepackage{bm}
\usepackage{romannum}
\usepackage{mathtools}
\usepackage{algorithm}
\usepackage{algpseudocode}
\usepackage{makecell}
\usepackage{tabularx,booktabs,caption,ragged2e}

\usepackage{tikzsymbols}

\def\etal{\textit{et al.}\xspace}

\def\eg{\textit{e.g.,}\xspace}

\def\aka{\textit{a.k.a.}\xspace}

\newcommand{\red}[1]{\textcolor{red}{#1}}

\author{Chen Chen}
\orcid{0000-0001-7179-0861}
\affiliation{%
  \department{Computer Science and Engineering}
  \institution{University of California San Diego}
  \city{La Jolla}
  \state{CA}
  \country{United States}
}
\email{chenchen@ucsd.edu}

\begin{document}

\title[From 2D Document Interactions into Immersive Information Experience]{From 2D Document Interactions into Immersive Information Experience: An Example-Based Design by Augmenting Content, Spatializing Placement, Enriching Long-Term Interactions, and Simplifying Content Creations}

\begin{teaserfigure}
    \includegraphics[width=\textwidth]{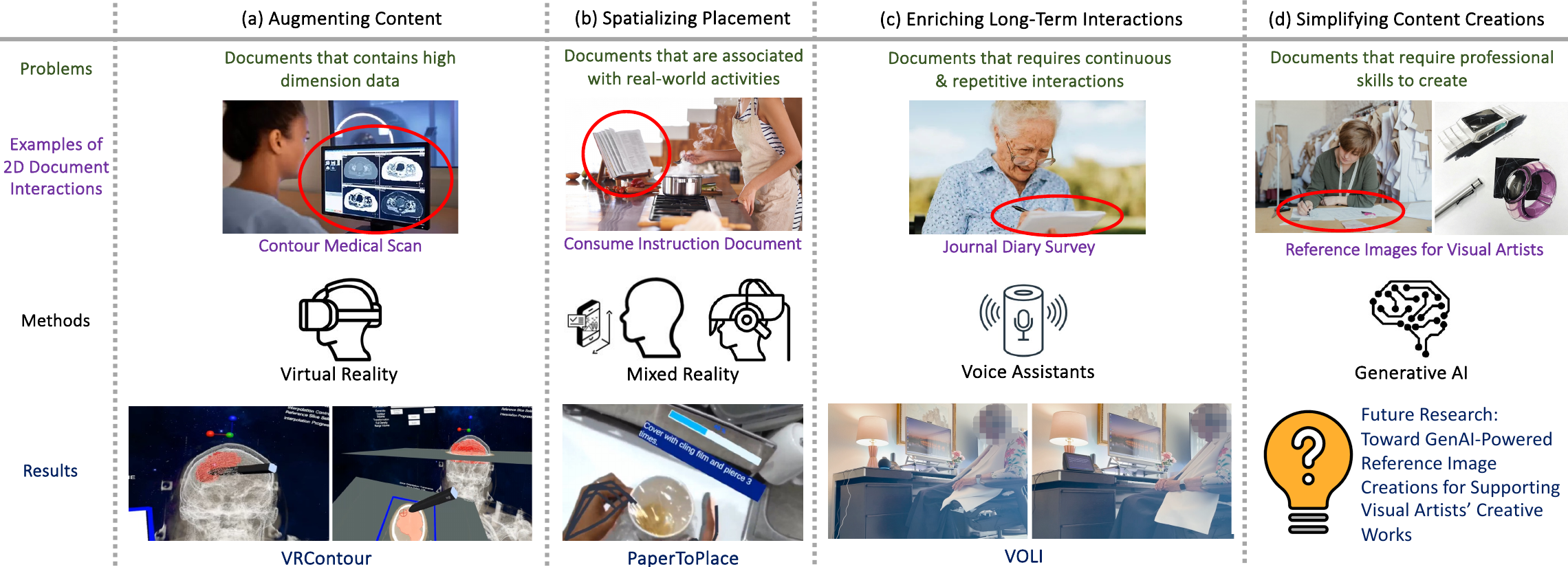}
    \caption{Research roadmap toward bringing and transforming today’s 2D document interactions into immersive information experience. The \red{red} circle highlighted the 2D documents in four examples.}
    \label{fig::teaser}
\end{teaserfigure}

\input{00-abstract}

\maketitle
\setcopyright{none}
\settopmatter{printacmref=false} 
\renewcommand\footnotetextcopyrightpermission[1]{} 
\pagestyle{plain}


\input{01-introduction}

\input{02-current}

\input{03-future}

\balance
\bibliographystyle{ACM-Reference-Format}
\bibliography{reference}

\end{document}

%% file: 00-abstract.tex
\begin{abstract}
{\it Documents} serve as a crucial and indispensable medium for everyday workplace tasks.
However, understanding, interacting and creating such documents on today's planar interfaces without any intelligent support are challenging due to our natural cognitive constraints on remembering, processing, understanding and interacting with these information.
My doctorate research investigates how to bring 2D document interactions into immersive information experience using multiple of today's emergent technologies.
With the examples of four specific types of documents --- medical scans, instruction document, self-report diary survey, and reference images for visual artists --- my research demonstrates how to transform such of today's 2D document interactions into an immersive information experience, by augmenting content with virtual reality, spatializing document placements with mixed reality, enriching long-term and continuous interactions with voice assistants, and simplify document creation workflow with generative AI.
\end{abstract}

%% file: 01-introduction.tex
\section{Introduction}\label{sec::introductions}
{\it Documents} serve as a crucial and indispensable medium for everyday workplace tasks.
For example, the instruction documents are commonly used to allow novice users to get acquainted with new tools and devices. 
Designers and visual artists use reference images to inspire, externalize and communicate their ideas.
In specialized professional settings such as healthcare, many professionals heavily rely on image-based information relayed by medical scans (\eg~CT and MRI) to understand diseases and proceed with diagnosis and treatment plans.

\begin{figure*}
    \centering
    \includegraphics[width=\linewidth]{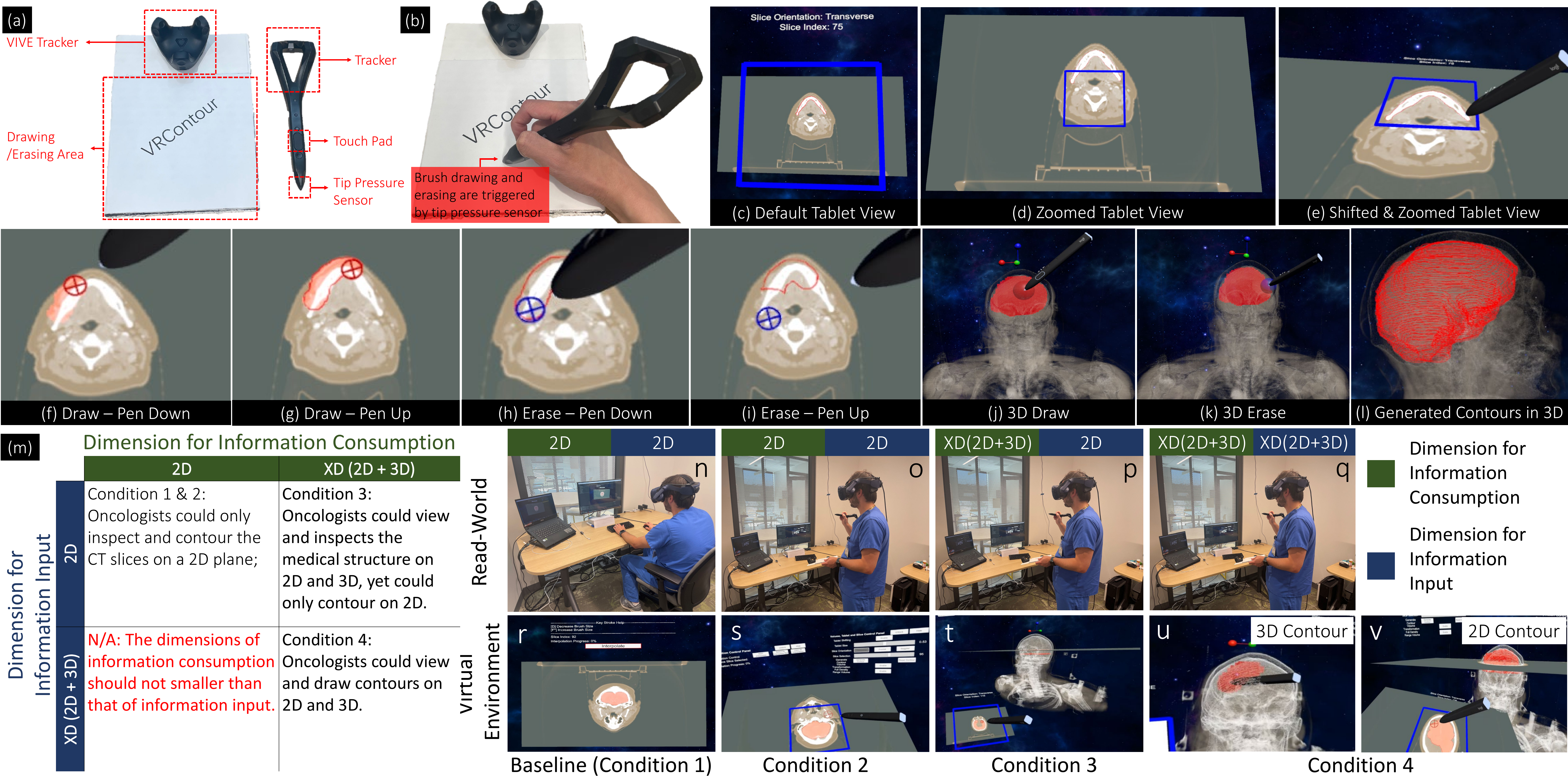}
    \vspace{-0.3in}
    \caption{VRContour~\cite{Chen2022VRContour}. (a - b) A VR stylus and a tracked tablet were used for supporting contour delineation workflow. (c - e) The virtual tablet rendered inside VR scene could be zoomed for supporting inspecting detailed structure. (f - l) The contour could be delineated and refined on 2D and 3D interfaces. (m) The design taxonomy of VRContour. (n - v) The final implementations of the VRContour.}
    \label{fig::vrcontour}
\end{figure*}

However, consuming and interacting with these paper-based documents without essential intelligence support can be challenging, due to our natural cognitive constraints on memorizing, processing, understanding and interacting with these information content~\cite{Atkinson1968}.
For example, while understanding medical scan documents that are 3D in nature, doctors need to go through each cross-sectional slices to build up mental models.
When following procedural steps in instruction documents for real-world tasks, novices must frequently switch attentions between the instructions and the tasks themselves.
Some documents such as self-report diary survey might require users to continuously interact with over the time. 
While this task may appear straightforward for younger individuals, it can pose significant challenges for older adults~\cite{Chen2021VOLINeeds}.
Despite the portability and enhanced document rendering capabilities of modern 2D computing displays and tablets, the fundamental challenges previously discussed remain prevalent.

The convergence of e\textbf{X}tended \textbf{R}eality (XR), \textbf{V}oice \textbf{A}ssistants (VAs), and \textbf{Gen}erative \textbf{AI} (GenAI) is ushering in an era of immersive information experiences, enabling us to rethink and explore different ways to create, deliver, consume and interact with existing 2D documents.
For example, the documents with high dimension data such as medical scans could be visualized by \textbf{V}irtual \textbf{R}eality (VR), which might be beneficial for doctors to interpret and understand the medical images more efficiently.
\textbf{M}ixed \textbf{R}eality (MR), in a similar way, unlocks an opportunity to computationally deliver and place the documents based on the real-world contexts.
This could be useful for the instruction documents that are usually required to be associated with real-world environment. 
VAs, on the other side, could be helpful for delivering documents that require users to interact continuously and repetitively over the time such as journaling self-report diary survey.
Finally, GenAI also leads to a promising research paradigm that enables efficient reference image creation workflow to support visual artists' creative works.
However, it is still unclear how 2D documents could be augmented with an immersive information experience using these emerging technologies.

While the concept of a {\it document} is defined as {\it ``a piece of written, printed, or electronic matter that provides information or evidence or that serves as an official record''}~\cite{documentDef}, but what will it look like after being brought into immersive experience?
Will it still be the {\it document}, as it is originally defined?
My research aims to tackle this problem with a user-centered design approach~\cite{Norman1986} contextualized on four different documents: the medical scans documents which contain high dimensional medical imaging data; the instruction documents which requires users to associate the content with real-world instructional activities; the self-report diary survey which requires individuals to access and interact continuously and repetitively; and the reference images for creative visual designs that requires professional drawing and images editing skills to create.
Figure~\ref{fig::teaser} demonstrates my research roadmap from high level.
To the end, we\footnote{While this abstract focuses on my Ph.D. research, the pronoun ``we'' would be used to recognize all the efforts from my collaborators.} demonstrated:

\begin{itemize}[noitemsep, topsep=0pt, leftmargin=*]

    \item how to {\it augment content dimension} to minimize overwhelm time and mental load while interacting with {\it high dimensional documents} such as a stack of medical scans; 

    \item how to {\it spatialize content placements} into workspace while interacting with {\it documents that are associated with real-world activities} such as instruction documents;

    \item how to {\it enable a long-term interactions} when it comes to the {\it documents that requires continuous and repetitive interactions} such as journaling diary;

    \item how to {\it facilitate an easy creation workflow} for {\it documents that need professional skills to create} such as reference images that could support visual artists' creative works\footnote{This line of research is considered as the future work, which will be addressed in the final year of my Ph.D.};

\end{itemize}

%% file: 02-current.tex
\section{VRContour: Boosting Contouring Experience by Augmenting Medical Scans with VR}\label{sec::past::vrcontour}
Interacting with medical scan documents is critical for radiotherapy treatment planning, and contouring is one indispensable step where oncologists need to identify and outline malignant tumors and/or healthy organs from a stack of medical images.
Inaccurate contouring could lead to systematic errors throughout the entire treatment course, leading to missing the tumors or over-treating the healthy tissues, and could cause increased risks of toxicity, tumor recurrence and even death~\cite{Zhai2021, Wuthrick2015}.

However, today’s contouring software such as Eclipse~\cite{Eclipse} and iContour~\cite{Yarmand2021iContourDesign21EA, Yarmand2023iContourCHI23EA, Yarmand2022iContourASTRO} are constrained to only work with a standalone 2D display, which is less intuitive and requires high task loads. 
Despite VR has shown great potentials in various specialties of healthcare and health sciences education due to the unique advantages of intuitive and natural interactions in immersive spaces, it has been unknown of how to bring contour delineation workflow into VR that requires capabilities to allow oncologists to inspect medical structures inside VR and precisely annotate on top of them.

VRContour~\cite{Chen2022VRContour} was the first effort that aimed to tackle this challenge with real-world medical professionals from UC San Diego Health System. 
Through our early work that quantitatively and qualitatively demonstrated how the commercially available VR input tools could better support the precision-first mid-air 3D drawing~\cite{Chen2022VRDrawing} and an autobiographical iterative design process with professional oncologists~\cite{Chen2022VRContourNeeds}, we first defined three design spaces focused on contouring in VR with the support of a tracked tablet and VR stylus.
We then designed and implemented the metaphors that the oncologists could delineate and refine the contours on 2D hand-held tablet (Fig.~\ref{fig::vrcontour}f - i), and 3D rendered volume (Fig.~\ref{fig::vrcontour}j - l), by considering today's mainstream contouring software with professional oncologists.
Fig.~\ref{fig::vrcontour}m lists four design spaces, including: inspecting and contouring planar medication structure contouring on 2D hand-held tablet; inspecting the 3D rendered medical structure and contour on the 2D tablet, where the delineated contours will be rendered into 3D; inspecting and contouring inside 3D medical structure and on 2D tablet (\aka~\textbf{C}ross-\textbf{D}imension (XD) contouring); and a baseline scenario to mock up current contouring workflow. 
Fig.~\ref{fig::vrcontour}n - v shows our implementation on HTC Vive Eye Pro with LogiTech VRInk being used as the VR stylus (see Fig.~\ref{fig::vrcontour}a - b).

Through a within-subject study with eight participants who have fundamental knowledge of basic anatomy yet without real-world contouring experience (\eg~senior M.D. students), we showed that visualizations of 3D medical structures could significantly increase precision (by nearly $60\%$, measured by Dice similarity coefficient~\cite{Zijdenbos1994}), and reduce mental load, frustration, as well as overall contouring effort. 
Participants also appreciated the benefits of using such metaphors for learning purposes.

\begin{figure*}[t]
    \centering
    \includegraphics[width=\linewidth]{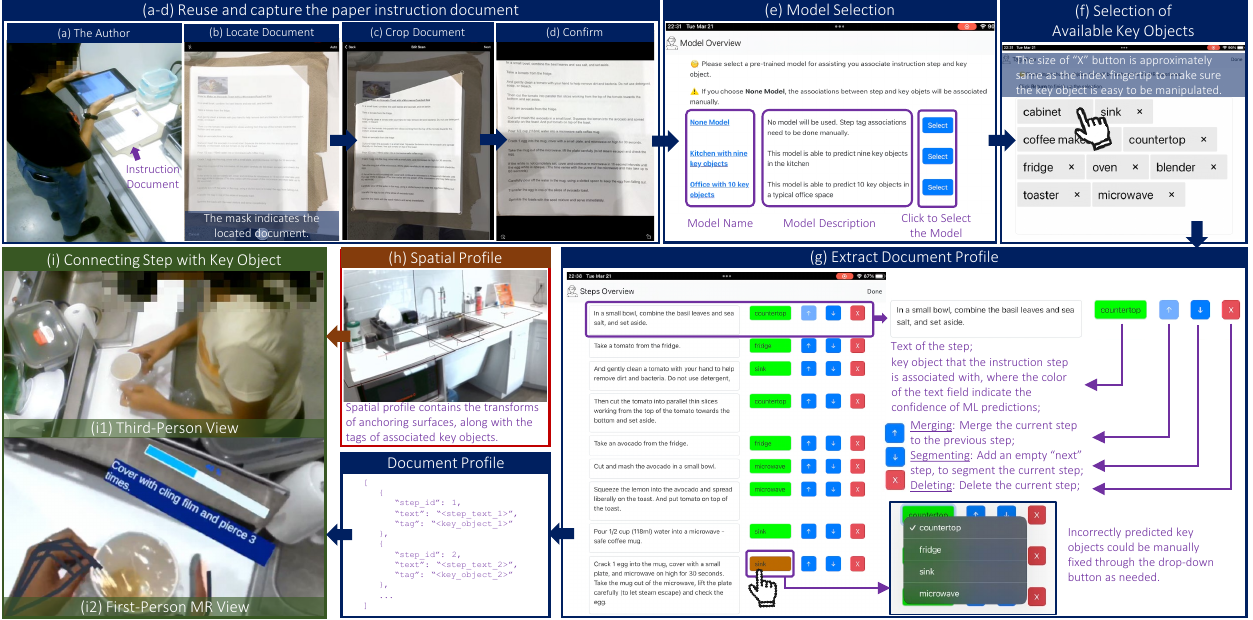}
    \caption{PaperToPlace~\cite{Chen2023PaperToPlace}. We assume a spatial profile has been pre-created (h). The author uses the authoring pipeline to extract the document profile for the MR experience (a - g). In the consuming pipeline (i), the instruction steps are displayed such that the consumers could easily refer to the instruction step while not being occluded by the virtual graphics.}
    \label{fig::paper2place}
\end{figure*}

\section{Paper-To-Place: Improving Instruction Consumption Experience by Spatializing Procedural Steps into Workspace with MR}\label{sec::past::papertoplace}
While many documents such as medical scans demonstrated in Sec.~\ref{sec::past::vrcontour} have 3D content in nature and could be easily augmented by using the unique affordances of spatial immersions brought by VR, many documents which do have such {``3D content''} might also need to be spatialized in the working environment.
Our work --- \mbox{PaperToPlace}~\cite{Chen2023PaperToPlace}, demonstrates how to bring everyday's paper-based instruction document into today's MR experience in a rapid, easier and context-aware approach.
While paper instructions are one of the mainstream medium for sharing knowledge, consuming such instructions and translating them into activities are inefficient due to the lack of connectivity with physical environment.
While tools, such as Microsoft Dynamic $365$ Guides~\cite{Dynamic365}, allow the instruction steps to be anchored in a predefined spatial location, real-world activities might change frequently, causing the virtual instructions might either block users' sight, or be too far to be read.

PaperToPlace~\cite{Chen2023PaperToPlace} (see Fig.~\ref{fig::paper2place}) demonstrates a novel workflow comprising an authoring pipeline, which allows the authors to rapidly transform and spatialize existing paper instructions into MR experience, and a consuming pipeline, which adaptive place each instruction step at an optimal location that is easy to read and do not occlude key interaction areas. 
Our evaluations of the authoring pipeline with $12$ participants demonstrated the usability of our workflow and the effectiveness of using a machine learning based approach to help extracting the spatial locations associated with each steps. 
A second within-subject study with another $12$ participants demonstrates the merits of our consumption pipeline by reducing efforts of context switching, delivering the segmented instruction steps and offering the hands-free affordances.

\begin{figure}[t]
    \centering
    \includegraphics[width=\linewidth]{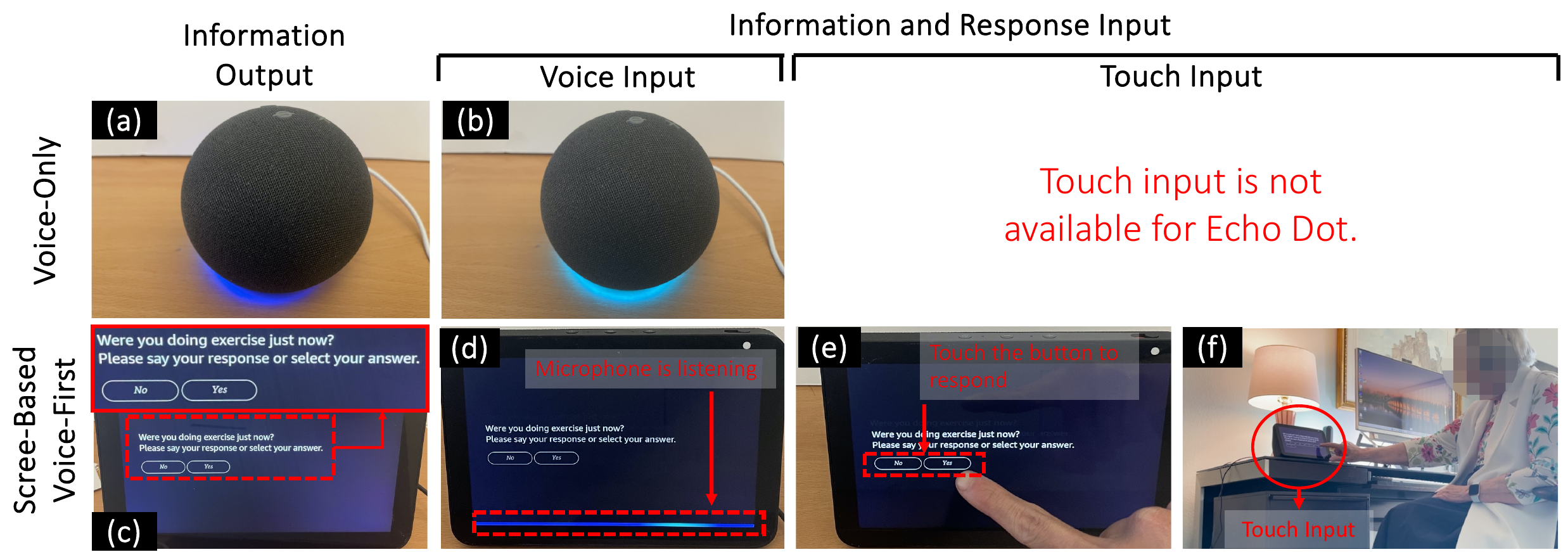}
    \vspace{-0.2in}
    \caption{We integrated the diary survey into Echo Dot and Echo Show, which were deployed into the residences of 16 older adults participants. The additional built-in touchscreen on Echo Show enables older adults to \emph{see} the survey prompt (c) and input responses by \emph{touch} (e - f).}
    \vspace{-0.20in}
    \label{fig::voli_deployment}
\end{figure}

\section{VOLI: Enriching Long-Term Interactions with Diary Survey among Aging Populations with VAs}\label{sec::past::voli}
More broadly, the immersive experience should not only be confined by visual immersion, having an {``always-on''}~\cite{Johnstone2022} auditory immersion is also critical for many document interactions. 
Leveraging such auditory enabler is also the indispensable key to push immersion toward temporal domain.
The self-report diary survey in today's geriatric practices, which need to be interacted {\it continuously} and {\it repetitively} is one type of document that urgely requires such transformation~\cite{Chen2021VOLINeeds, Chen2021VOLIDeployment, Lifset2023}.

Understanding older adults' physical and mental states are important in today's geriatric practices. 
Such diary data are typically collected by phone calls from triage nurses, which might incur additional costs and complexities, or requiring older adults to self report through a web-based patient portal, which might be challenging for those without proficient computing experience~\cite{Chen2021VOLINeeds}.
While voice user interfaces offer increased accessibility due to hands-free and eyes-free interactions, older adults often have challenges such as constructing structured requests and perceiving how such devices operate. 

With emphasis on privacy~\cite{Sun2020}, one thread of my Ph.D. research focused on the voice-first user interfaces which are promising to address these challenges by enabling multimodal interactions~\cite{voli}. 
Standalone voice $+$ touchscreen VAs, such as Echo Show, are specific types of devices that adopt such interfaces and are gaining popularity. 
However, the affordances of the additional touchscreen for older adults were still unknown. 
My research integrated the self-report diary survey into the Echo Dot and the Echo Show --- the dominant and representative voice-only and voice-first screen-based voice assistants (see Fig.~\ref{fig::voli_deployment})~\cite{Chen2021CUI}.
We then conducted a first within-subjects study over $40$-day real-world deployment with $16$ older adults with average age of $82.5$ ($SD = 7.77$) to understand how a built-in touchscreen might benefit older adults' experience while conducting self-report diary survey.

Our results~\cite{Chen2021VOLIDeployment} showed that the capabilities of visualizing diary document content through touchscreen is useful to enhance diary compliance.
The modality of touch input could also reduce the response latency, even though the older adults still preferred to journal diary data through speech. 
The insights generated through this deployment study offers indispensable guidance to immerse diary interactions into older adults' life through voice-first VAs.

%% file: 03-future.tex
\section{Conclusions and Future Work: Toward GenAI-Powered Reference Images Creations for Supporting Visual Artists' Creative Works}
Interacting with documents involves more than just reading and annotating them, the {\it creation} process is equally important.
The final line of my Ph.D. research {\it will} focus on the creations of the reference images for creative design workflow.
In nearly all creative visual design process, reference images are considered as an important and indispensable medium for inspiring~\cite{Mougenot2008} as well as externalising and communicating~\cite{Kang2018, Herring2009} ideas.
However, creating reference images could be challenging, as it requires designers to have professional image sketching and editing skills.
Although many designers might simply create reference images by searching internet images, these searched images might not be able to perfectly grasp and externalize designers' thoughts, leading to communicating incorrect design gist.
In many cases, the internet searched images might be created by similar types of content creators, which might pose ``design fixations'' when designers are attempting to distill and transfer the gist of the reference images to their own design~\cite{Jansson1991}. 

Today's \textbf{L}arge-scale \textbf{T}ext-to-image \textbf{G}eneration \textbf{M}odels (LTGMs) trained on huge dataset, such as the pretrained stable diffusion based text-to-image model~\cite{Rombach2022}, has demonstrated the capabilities for creating high-quality open-domain images from textual prompts.
These LTGMs have showed the potentials to support visual artists' creative works due to their capabilities to create anthropomorphized versions of objects and animals, combine irrelevant concepts in reasonable ways, or even generate variations of the additional input images~\cite{Ko2023, Son2023}.
However, through an extensive literature and interview study, Ko~\etal~\cite{Ko2023} identified four key setbacks, including lacks of support for different types of visual arts; requirement of model customization grounded on the domain-specific understanding; needs of more control of synthesized images; and assistance for crafting and optimizing textual prompts.
These setbacks also pose challenges on designing creativity support tools for generating reference images by leveraging the power of GenAI. 

The last thread of my doctorate research aims to design a creativity support tools to help designers create visual reference images using GenAI, to facilitate a more efficient inspiration and communications.
We will conduct a formative study with professional designers, followed by prototyping a demonstrable working system for realizing such vision.
A real-world user study will also be conducted in the final stage.
During the symposium, I will discuss my past projects, the potentials of future direction and long-term vision.